\documentclass[useAMS,usenatbib]{mnras}
\usepackage{amsmath}
\usepackage{amssymb}
\usepackage{graphicx}
\usepackage{subfig}
\usepackage{url}

\title{Probing the Gas Density in our Galactic Center: Moving Mesh Simulations of G2}
\author[E. Steinberg etl al.]{Elad Steinberg$^1$\thanks{E-mail: elad.steinberg@mail.huji.ac.il}, Re'em Sari$^1$, Orly Gnat$^1$,
	Stefan Gillessen$^2$, Philipp Plewa$^2$,\newauthor
	Reinhard Genzel$^2$, Frank Eisenhauer$^2$, Thomas Ott$^2$, Oliver Pfuhl$^2$, Maryam Habibi$^2$, \newauthor
	Idel Waisberg$^2$, Sebastiano von Fellenberg$^2$, Jason Dexter$^2$, Michi Baub\"{o}ck$^2$ \newauthor and Alejandra Jimenez Rosales$^2$\\
	$^1$ Racah Institute of Physics, Hebrew University, Jerusalem 91904, Israel\\
	$^2$ Max-Planck-Institut für Extraterrestrische Physik, D-85748 Garching, Germany}
\begin{document}
	\maketitle
\input{aas_macros.sty}
\begin{abstract}
The G2 object has recently passed its pericenter passage in our Galactic Center. While the $Br_\gamma$ emission shows clear signs of tidal interaction, the change in the observed luminosity is only of about a factor of 2, in contention with all previous predictions.
We present high resolution simulations performed with the moving mesh code, RICH, together with simple analytical arguments that reproduce the observed $Br_\gamma$ emission. In our model, G2 is a gas cloud that undergoes tidal disruption in a dilute ambient medium. We find that during pericenter passage, the efficient cooling of the cloud results in a vertical collapse, compressing the cloud by a factor of $\sim5000$. By properly taking into account the ionization state of the gas, we find that the cloud is UV starved and are able to reproduce the observed $Br_\gamma$ luminosity. 
For densities larger than $\approx500\;\mathrm{cm}^{-3}$ at pericenter, the cloud fragments, due to cooling instabilities and the emitted radiation is inconsistent with observations. For lower densities, the cloud survives the pericenter passage intact and its emitted radiation matches the observed lightcurve.
From the duration of $Br_\gamma$ emission which contains both redshifted and blueshifted components, we show that the cloud is not spherical but rather elongated with a size ratio of 4 at year 2001. The simulated cloud's elongation grows as it travels towards pericenter and is consistent with observations, due to viewing angles. The simulation is also consistent with having a spherical shape at apocenter.
\end{abstract}
\section{Introduction}
Half a decade since its initial discovery \citep{G2}, the exact nature of the G2 object and its future evolution remains a mystery. Although there has been a large observational campaign to observe the cloud in various wavelengths \citep{G2,Phifer,Witzel,Spitzer,microwave,radio,Pfuhl,Plewa}, so far, no model has managed to match the observations.
Broadly speaking, there are two classes of models for the G2 object. 
According to the first, G2 is a gas cloud with a mass of a few earth masses that may or may not have a stellar object inside \citep{G2,gas_star,Witzel,James}. 
The second class suggests G2 is composed of gas being actively ejected from the vicinity of a central object \citep{avi,wind}.
Numerous simulations have been carried out to date, simulating both in 2D and in 3D the gas cloud scenario \citep{Schartmann2012,moving,sph,Schartmann2015} as well the ejected gas model \citep{wind2d,wind3d2,wind3d}. 

Until now, no model has successfully explained the observed $Br_\gamma$ emission during pericenter passage, either predicting a large increase in the emission or predicting a drastic reduction, neither of which are consistent with the relatively flat lightcurve.
Emission in the radio has also been predicted that results from either a bow shock that is formed at the head of the cloud as it passes through the ambient medium or from the bow shock where the ejected wind interacts with the ambient medium \citep{radio_emit,radio_emit2}. However, no significant change in the radio flux from the Galactic Center has been observed during the pericenter passage of the cloud \citep{radio}.

In this paper we present a detailed model for the evolution and $Br_\gamma$ emission of the gas cloud model that combines analytical insights aided by comprehensive numerical simulations carried out with the RICH code \citep{Rich}.
Our numerical simulations include:
\begin{itemize}
	\item Extremely high spatial resolution enabled by the unstructured semi-Lagrangian nature of RICH.
	\item A novel method of combining two 2D simulations to achieve an approximate 3D picture, thus enabling a very high ``effective" 3D resolution that is unattainable otherwise.
	\item Proper handling of the gas thermodynamics by actively cooling the gas instead of assuming an isothermal equation of state.
	\item Self consistent calculation of the $Br_\gamma$ emission by taking into account the ionization state of the gas.
\end{itemize}

In section \ref{sec:pre} we investigate the dynamics of the gas cloud until it reaches pericenter. The dynamics of the pericenter passage are given in section \ref{sec:pericenter}. Post-pericenter evolution is presented in section \ref{sec:post}. In section \ref{sec:size} we present a robust estimate of the size of the cloud along the orbit derived from the simultaneous emission of redshifted and blueshifted lines. Our numerical simulations and their results are shown in section \ref{sec:hydro}, while the $Br_\gamma$ emission that is calculated from them is presented in section \ref{sec:brg}. In section \ref{sec:discuss} we discuss the implication of our results and summarize our work.
\section{Pre-Pericenter Dynamics in the Orbital Plane}
\label{sec:pre}
The full dynamical and physical evolution of the G2 gas cloud is an intrinsically 3D problem. However, since we run two 2D planar simulations, we explain below analytically, how the dynamics and physical state of the cloud evolve in 2 and 3 dimensions.
\subsection{2D}
\begin{figure}
	\centering
	\includegraphics[width=0.95\linewidth]{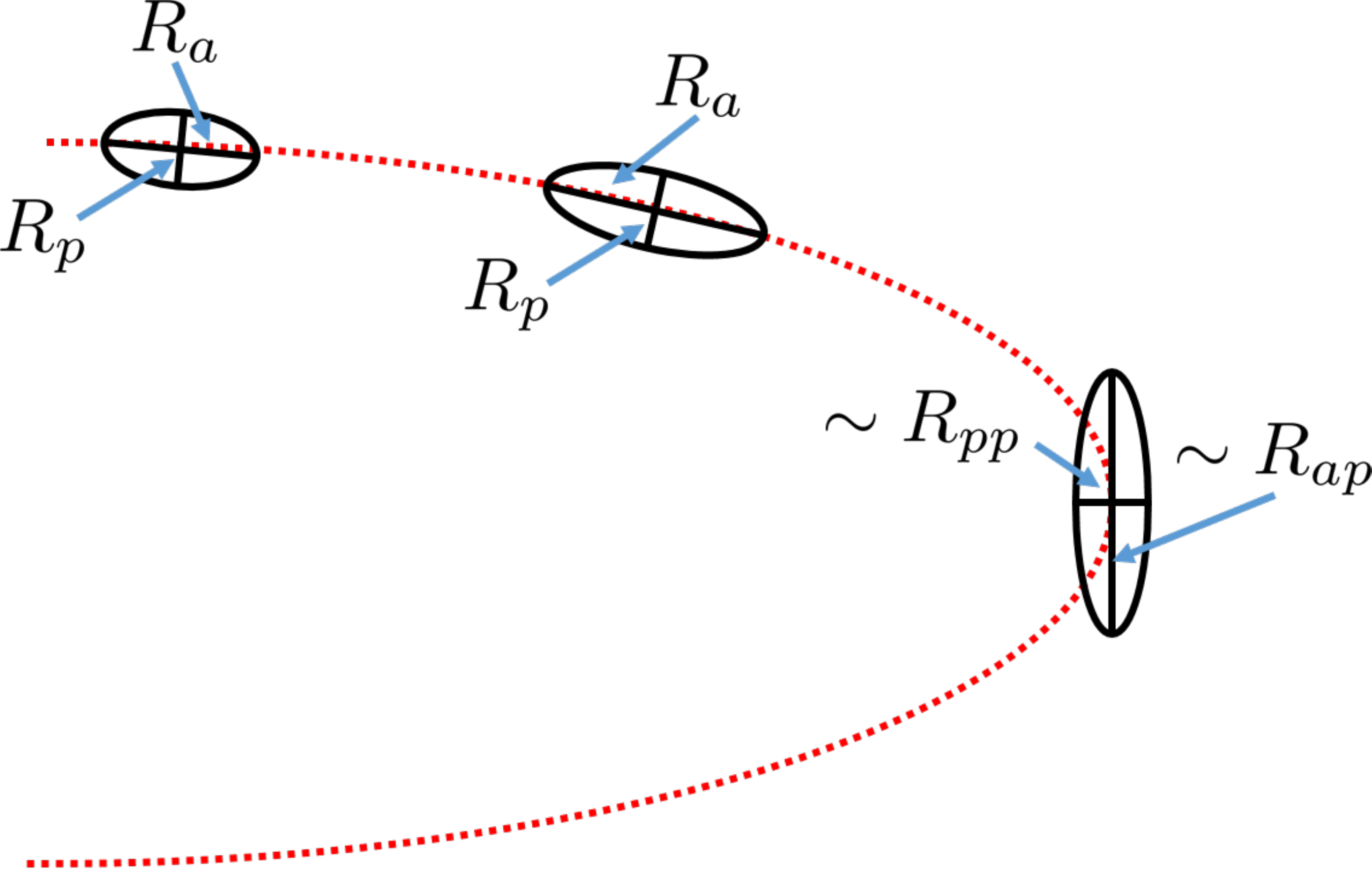}
	\caption{A schematic cartoon of the length of G2 along (perpendicular) the orbit as described in eq. \ref{eq:R}.}
	\label{fig:R}
\end{figure}

During its infall, the cloud is stretched along the orbit and squeezed in the direction perpendicular to the orbit due to tides. Approximating its orbit to be parabolic far from pericenter, \cite{HVS} have shown that the length of the cloud along (perpendicular)  the orbit, $R_a\;(R_p)$, evolves as
\begin{eqnarray}
\label{eq:R}
R_{a}&=&R_{ap}\sqrt{\frac{r_p}{r}}\\
R_{p}&=&R_{pp}\sqrt{\frac{r}{r_p}}
\end{eqnarray}
where $r_p$ is the pericenter distance, $R_{ap}$ is the length of the cloud along the orbit at pericenter and $R_{pp}$ is the length of the cloud perpendicular to the orbit at pericenter. Far from pericenter this approximation is accurate, while at pericenter the values of $R_a$ and $R_p$ are correct up to a factor of order unity. In the 3D scenario that is discussed in sec.\ref{sec:3D}, the value of $R_p$ perpendicular to the orbital plane is inaccurate at pericenter since the height of the cloud goes to zero. Figure \ref{fig:R} shows a cartoon of the various lengths with respect to the orbit.

Since in 2D the area of the cloud is roughly conserved, the pressure, $P$, and density, $\rho$, of the cloud are approximately constant.

Tidal forces accelerate the gas in the direction perpendicular to the orbit as it travels towards pericenter. Assuming a SMBH of mass $M=4.3\cdot 10^6 M_\odot$,  the perpendicular velocity at pericenter due to tides is:
\begin{equation}
\begin{split}
	\dot{R}^{tide}_{p}&\approx R_{p}\Omega/\sqrt{2}\approx 3.6\cdot 10^7\left(\frac{2r_p}{r}\right)\left(\frac{R_{pp}}{3\cdot 10^{14}\;\mathrm{cm}}\right)\\
	&\times\left(\frac{1.75\cdot 10^{15}\;\mathrm{cm}}{r_p}\right)^{3/2}\;\mathrm{cm/s}
\end{split}
\label{eq:tide_vel}
\end{equation}
where $\Omega$ is the orbital frequency. This compression does not give rise to shock wave formation due to its homologous nature.

Once the cloud travels into a region where the pressure is greater than its own, a shock wave that compresses the cloud forms at the outer edges of the cloud.
Assuming the ambient atmosphere has a profile of $P_{atm}\sim r^{-2}$, the cloud gets compressed at a rate of:
\begin{equation}
\begin{split}
		\dot{R}^{atm}_{p}&\approx \sqrt{\frac{P_{atm}}{\rho}}\approx c\sqrt{\eta}\frac{r_p}{r}\approx 1.6\cdot 10^7\left(\frac{2r_p}{r}\right)\\
		&\times\left(\frac{\eta}{1000}\right)^{1/2}\;\mathrm{cm/s}
\end{split}
\label{eq:shock}
\end{equation}
assuming $c$, the sound speed of an isothermal cloud, corresponds to a temperature of $10^4$ K (the temperature at which the cooling timescale becomes longer than the orbital timescale). $\eta$ is the ratio between the ambient atmosphere's pressure and the cloud's pressure at pericenter assuming a radial orbit until pericenter. In 2D, the actual ratio between the pressures is different by a factor of order unity relative to $\eta$ since the orbit is parabolic and not radial; while in 3D, it can be very different due to the compression in the cloud's vertical axis.

In reality, the compression in the perpendicular direction has contributions from both the compression by tides as well as the compression from the ambient atmosphere. However, the result is less than the naive sum of the two terms mentioned above, since once the cloud shrinks in the perpendicular direction due to the ambient gas it would reduce the compression due to the tides. 

The width of the area that will be shock heated is given by the shock velocity, $\dot{R}^{atm}_{p}$ times the cooling time, $\frac{P_{atm}}{\Lambda(T) n_H^2}$:
\begin{equation}
\Delta R \approx \frac{P_{atm}\dot{R}^{atm}_{p}}{\Lambda(T) n_H^2}\approx \frac{\rho c^3}{\Lambda(T) n_H^2}\eta^{3/2}\left(\frac{r_p}{r}\right)^3\label{eq:shock_width}
\end{equation}
where $\Lambda(T)$ is the cooling function for optically thin gas and $n_H$ is the post shock hydrogen number density.

The shock temperature is given by:
\begin{equation}
	T_s\approx 2.5\cdot 10^6\left(\frac{\eta}{1000}\right)\left(\frac{2r_p}{r}\right)^2\;\mathrm{K}\label{eq:shock_T}.
\end{equation}
Combining eq. \ref{eq:shock_T} and eq. \ref{eq:shock_width} gives:
\begin{equation}
\begin{split}
\Delta R&\approx3.75\cdot 10^{13}\left(\frac{\eta}{1000}\right)^{3/2} \left(\frac{3\cdot 10^{-19}\;\mathrm{g/cm^3}}{\rho}\right)\left(\frac{c}{10^6\;\mathrm{cm/s}}\right)^3\\
&\times\left(\frac{5.6\cdot 10^{-23}\;\mathrm{erg\;cm^3\;s^{-1}}}{\Lambda(T)}\right)\left(\frac{2r_p}{r}\right)^3\;\mathrm{cm}.
\end{split}
\end{equation}
There is no adiabatic heating due to compression by tides since the cloud does not change its area.
\subsection{3D}
\label{sec:3D}
Taking into account the full 3D picture, we also have compression along the $z$ axis (the axis perpendicular to the orbital plane), which behaves similarly to $R_{p}$ until pericenter passage. 

Assuming the cloud is isothermal, the pressure and density scale as:
\begin{equation}
	P\sim\rho\sim \sqrt{\frac{r_p}{r}}.
\end{equation}

The compression due to the ambient atmosphere is:
\begin{equation}
\begin{split}
	\dot{R}^{atm}_{p}&\approx \sqrt{\frac{P_{atm}}{\rho}}\approx c\sqrt{\eta}\left(\frac{r_p}{r}\right)^{3/4}\approx 1.9\cdot 10^7\left(\frac{\eta}{1000}\right)^{1/2}\\
	&\times\left(\frac{2r_p}{r}\right)^{3/4}\;\mathrm{cm/s}.
\end{split}
\end{equation}

The shock temperature is given by:
\begin{equation}
T_s\approx 3.5\cdot10^6\left(\frac{\eta}{1000}\right)\left(\frac{2r_p}{r}\right)^{3/2}\;\mathrm{K}\label{eq:shock_T3D}.
\end{equation}
Combining eq. \ref{eq:shock_T3D} and eq. \ref{eq:shock_width} gives us the width of the hot shell:
\begin{equation}
\begin{split}
\Delta R & \approx2.7\cdot 10^{13}\left(\frac{\eta}{1000}\right)^{3/2}\left(\frac{1.5\cdot 10^{-18}\;\mathrm{g/cm^3}}{\rho(r=r_p)}\right)\\
&\times\left(\frac{c}{10^6\;\mathrm{cm/s}}\right)^3\left(\frac{3.7\cdot 10^{-23}\;\mathrm{erg\;cm^3\;s^{-1}}}{\Lambda(T)}\right)\left(\frac{2r_p}{r}\right)^{1.75} \;\mathrm{cm}.
\end{split}
\end{equation}

The cloud remains isothermal since the cooling timescale is much shorter than the adiabatic heating timescale. The ratio of timescales is:

\begin{equation}
\begin{split}
\frac{\tau_{adiabatic}}{\tau_{cool}}&\approx\frac{\sqrt{2}\Lambda\rho}{c^2m_p^2\Omega}\approx 30\left(\frac{r}{2r_p}\right)^2\left(\frac{10^6\;\mathrm{cm/s}}{c}\right)^2\\ &\times\left(\frac{\Lambda}{3\cdot10^{-24}\;\mathrm{erg\;cm^3\;s^{-1}}}\right)\\
&\times\left(\frac{\rho(r=r_p)}{1.5\cdot 10^{-18}\;\mathrm{g/cm^3}}\right)\left(\frac{r_p}{1.75\cdot10^{15}\;\mathrm{cm}}\right)^2
\end{split}
\end{equation}
where $m_p$ is the mass of the proton.
To summarize, tides stretch the cloud along the orbit and compress it perpendicular to the orbit. During its infall, most of the cloud remains isothermal with a thin shell of shocked gas at its outer edges.
\section{Pericenter Passage}
\label{sec:pericenter}
During pericenter passage, the ambient pressure and compression velocity are approximately constant.
\subsection{2D}
The cloud is compressed perpendicular to the orbit by the ambient atmosphere until it reaches the pressure of the atmosphere during pericenter passage.
The ratio between the timescale for the cloud to reach pressure equilibrium with the ambient atmosphere and the Keplerian timescale is given by:
\begin{equation}
\begin{split}
	t_{eq}/t_{Keplerain}&\approx \frac{\Omega R_p}{\pi\dot{R}^{atm}_{p}}\approx\Omega(r=r_p) \eta^{-1/2}\frac{R_{pp}}{\pi c}\\
	&\approx 0.7 \left(\frac{1000}{\eta}\right)^{1/2}\left(\frac{R_{pp}}{3\cdot 10^{14}\;\mathrm{cm}}\right)^{3/2}.
\end{split}
\end{equation}
For $\eta\approx1000$ the cloud marginally reaches pressure equilibrium with the ambient atmosphere, while for lower $\eta$ it does not reach pressure equilibrium.

In the process of achieving pressure equilibrium, the cloud undergoes shock heating and rapid cooling that might lead to cooling instabilities that fragment the cloud into dense clumps.

\subsection{3D}
The compression velocity in the vertical axis is roughly constant throughout pericenter passage and the heating timescale due to tides is given by:
\begin{equation}
\tau_{heat}\approx\frac{R_{p}}{\dot{R}_{p}^{tide}}.
\end{equation}
Since at pericenter the timescale for cooling is shorter than for the heating (by about a factor of 7), and since they scale the same way with regards to the density of the cloud, the cloud remains isothermal. 

During pericenter passage tides do not change the area of the cloud in the orbital plane but the collapse in the vertical direction continues.
Stopping the collapse of the cloud, requires it to have a sound speed that is comparable to $\dot{R}_{p}^{tide}$. 
Since the cloud can cool efficiently and is roughly isothermal, naively, the compression can not be stopped and the cloud can be infinitely compressed.
However, the shock wave that is induced by the ambient atmosphere breaks the homologous nature of the collapse. Fluid elements behind the shock front collapse with a velocity greater than the homologous velocity of the tidal field.
The stronger the shock, the faster the shock front reaches the orbital plane and hence the cloud has less time for a homologous collapse. Ambient atmospheres with a lower pressure lead to a stronger collapse and vice versa.

\section{Post-Pericenter}
\label{sec:post}
Assuming the cloud undergoes a cooling instability, two phases of gas arise. The first is composed of cold dense clumps that manage to reach pressure equilibrium by having their original density increased by a factor $\eta$.

The second phase is shock heated to $\approx T_s$ while having an increase in density of order a few. The cooling time for the latter phase (assuming $\eta=1000$) is  :
\begin{equation}
t_{cool}\approx 10^9\;\mathrm{s}.
\end{equation}
If the cloud does not undergo a cooling instability, its size in the orbital plane is roughly constant while it expands in the vertical direction with a velocity comparable to its sound speed.
\section{Size Constraints from Observations}
\label{sec:size}
The observed PV diagrams \citep{Pfuhl,Plewa} show that there is simultaneous emission from the redshifted and the blueshifted parts of the cloud for a period of order $\Delta t =1-2$ years. Both the length of the cloud along the orbit and its Keplerian velocity scale the same way with regards to the distance from the SMBH, implying that the time it takes the cloud to travel its own size is roughly constant. This suggests (assuming that the extent of the cloud is larger than the pericenter distance) that the extent of the cloud along the orbit is
\begin{equation}
R_{ap}\approx \sqrt{2}\Omega(r=r_p)r_p\Delta t\approx
2.5-5\cdot10^{16}\sqrt{\frac{1.75\cdot10^{15}\;\mathrm{cm}}{r_p}}\;\mathrm{cm}.
\label{eq:Rap}
\end{equation}
While this size is larger than the current observational PSF by an order of magnitude, the orbit's orientation is such that the direction along the orbit is close to the line of sight (with the exception of the pericenter passage), and thus only a fraction of this elongation has been observed.

\section{Hydrodynamical Simulations}
\label{sec:hydro}
Accurately capturing the dynamics and shock front in a full 3D simulation requires extreme resolution that is currently unachievable \citep{Schartmann2015}. Instead, we perform two 2D simulations that independently evolve the dynamics in the orbital plane and the vertical collapse\footnote{Movies of the simulations are available at \url{https://www.eladsteinberg.com/movies}}.

\begin{figure}
\centering
\includegraphics[width=0.9\linewidth]{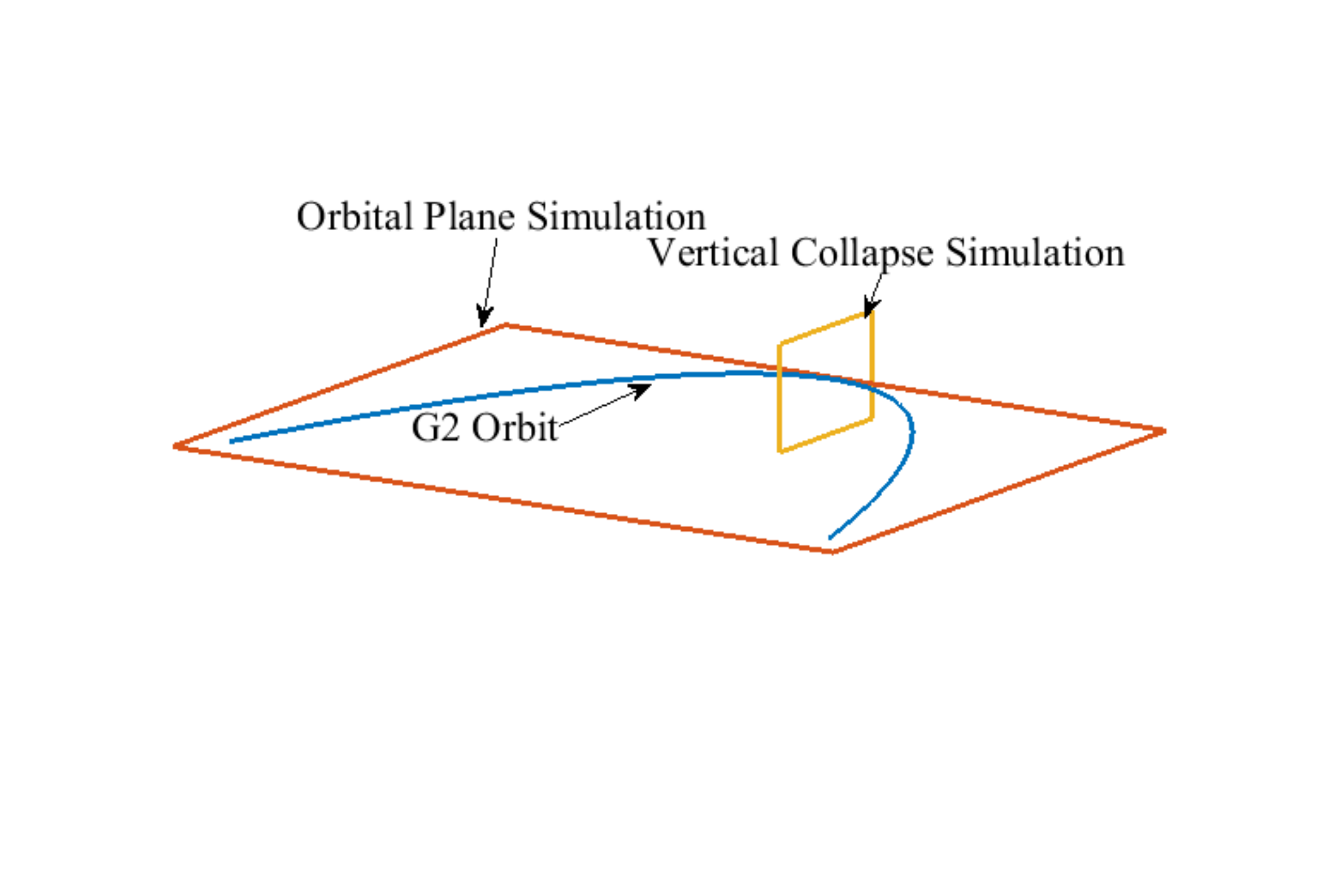}
\caption{An illustration of the two 2D simulations we perform. The first simulates the dynamics in the orbital plane of G2 and the second simulates the vertical collapse of G2 as it passes through pericenter.}
\label{fig:simulations}
\end{figure}

An illustration of our two setups is shown in fig. \ref{fig:simulations}.

We run our simulations using the RICH code \citep{Rich} which is a publicly available Voronoi based finite volume code. RICH has the benefit of accurately capturing shocks and discontinuities as any Eulerian Godunov scheme while minimizing the advection term since it is a semi-Lagrangian code. 

Initially we set the ambient atmosphere to be the RIAF solution as presented in \cite{Yuan2003}
\begin{eqnarray}
n_{atm}&=&1654\left(\frac{10^{16}\mathrm{cm}}{r}\right)\mathrm{cm}^{-3}\\
T_{atm}&=&2.13\cdot 10^8\left(\frac{10^{16}\mathrm{cm}}{r}\right)\mathrm{K}
\label{eq:atm}
\end{eqnarray}
where $n_{atm}$ is the number density, $T_{atm}$ is the temperature and we have adopted a mean molecular weight of $\mu=0.6139$. This profile arises from a fit to $Chandra$ observations of the SMBH atmosphere done at a distance of $\approx0.1$ pc from the SMBH, which is extrapolated towards the Galactic Center and corresponds to a value of $\eta\approx800$. As stated in \cite{Schartmann2012} this profile is convectively unstable, and as a consequence we employ the same method as in \cite{Schartmann2012} to stabilize it. Specifically, we advect a passive tracer that traces the mass fraction in a cell that originates from the cloud. If the mass fraction is less than $10^{-6}$ we reset the cell's primitive variables to be the ambient atmosphere values at the end of each time step.
\subsection{Simulation in the Orbital Plane}
Our coordinate system is centered around the SMBH, the $x$ axis points towards pericenter and the $y$ axis is perpendicular to it.
The gravity of the SMBH is represented by a point source term and we assume a SMBH mass of $4.3\cdot10^6M_\odot$.
In order to match the observed $Br_\gamma$ emission at early times as well as match the duration of pericenter passage, we set up the cloud as a homogeneous ellipse with a minor axis of $R_c=1.87\cdot10^{15}\mathrm{cm}$ and a major axis of $4R_c$  aligned along the Keplerian orbit. The density is assumed to be $\rho_c=3.1\cdot10^{-19}\;\mathrm{g/cm^3}$. 

These initial conditions are consistent both with the size constraint given in section \ref{sec:size} and with the measured $Br_\gamma$ luminosity and resolved size given in \cite{G2}. The temperature is assumed to be $T_c=10^4\mathrm{K}$ and the center of the cloud is located at $\vec{r}_0=(-6.74\cdot10^{16},\; 1.89\cdot10^{16})$ cm, corresponding to the year 2001.
The cloud's initial velocity corresponds to a Keplerian orbit with a pericenter distance of $r_p=1.7\cdot10^{15}$ cm and an eccentricity of $e=0.987$, based on early estimates from the latest set of observations. This orbit is very similar to the most recent result \citep{Plewa} and its pericenter differs by only $17\%$.

The box size for our simulations is $[-7.55,-2.25]\times[1.89,2.25]\cdot 10^{16}$ cm, and the boundary conditions are set to allow matter to flow out but not in. We impose an inner boundary at a radius of $2.43\cdot 10^{14}$ cm from the SMBH where matter can flow inside it but not out. An adiabatic equation of state of ideal gas is assumed with an adiabatic index of $\gamma=5/3$. We use approximately $4\cdot10^5$ cells, most of which contain gas from the cloud giving us a maximum resolution of $7\cdot10^{12} $ cm during pericenter passage. In the inner $10^{15}$ cm we split cells whose size is larger than $5\cdot10^{13}$ cm.

\begin{figure*}
	\centering
	\subfloat{{\includegraphics[width=0.5\linewidth]{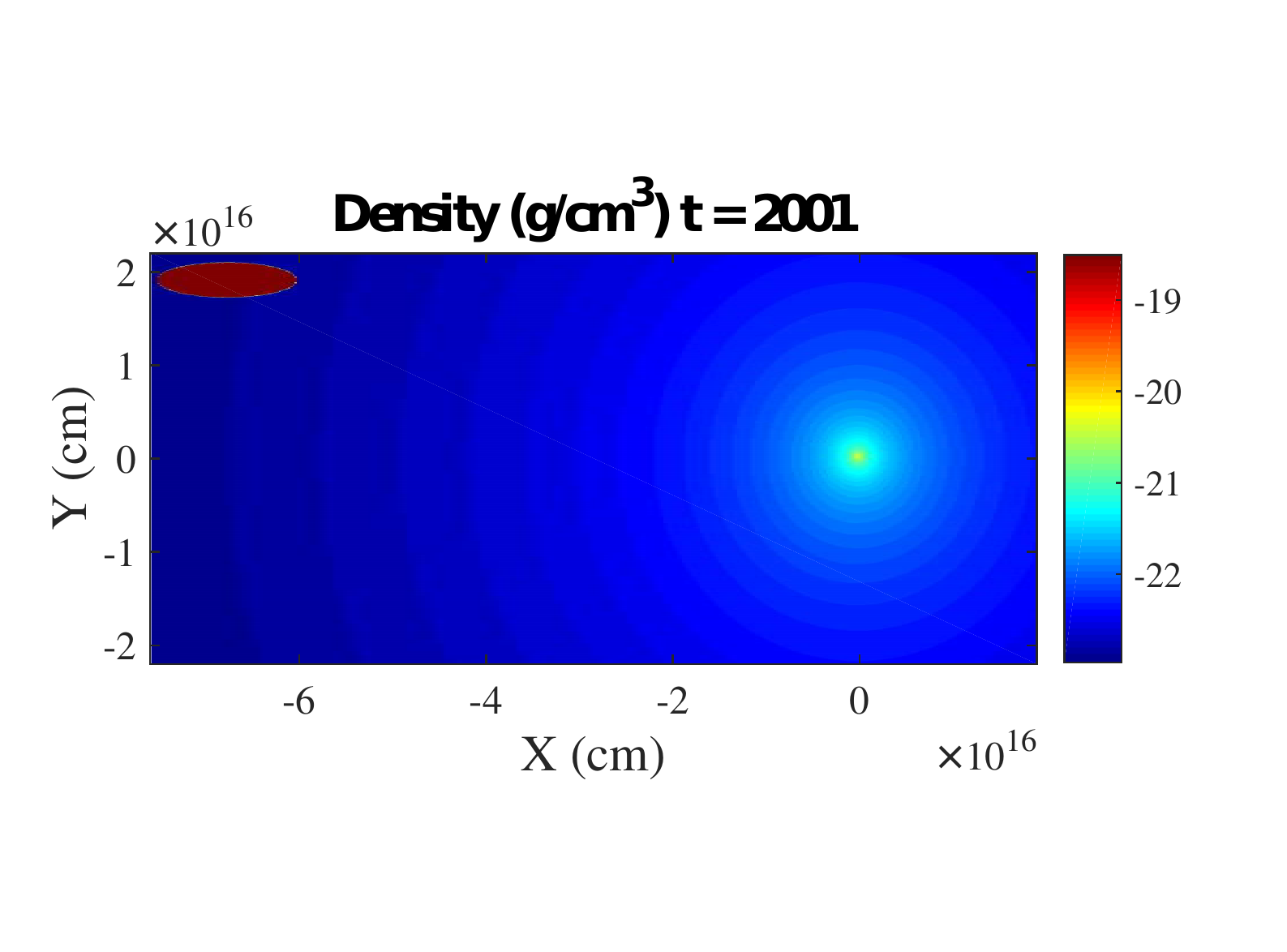}}}
	\subfloat{{\includegraphics[width=0.5\linewidth]{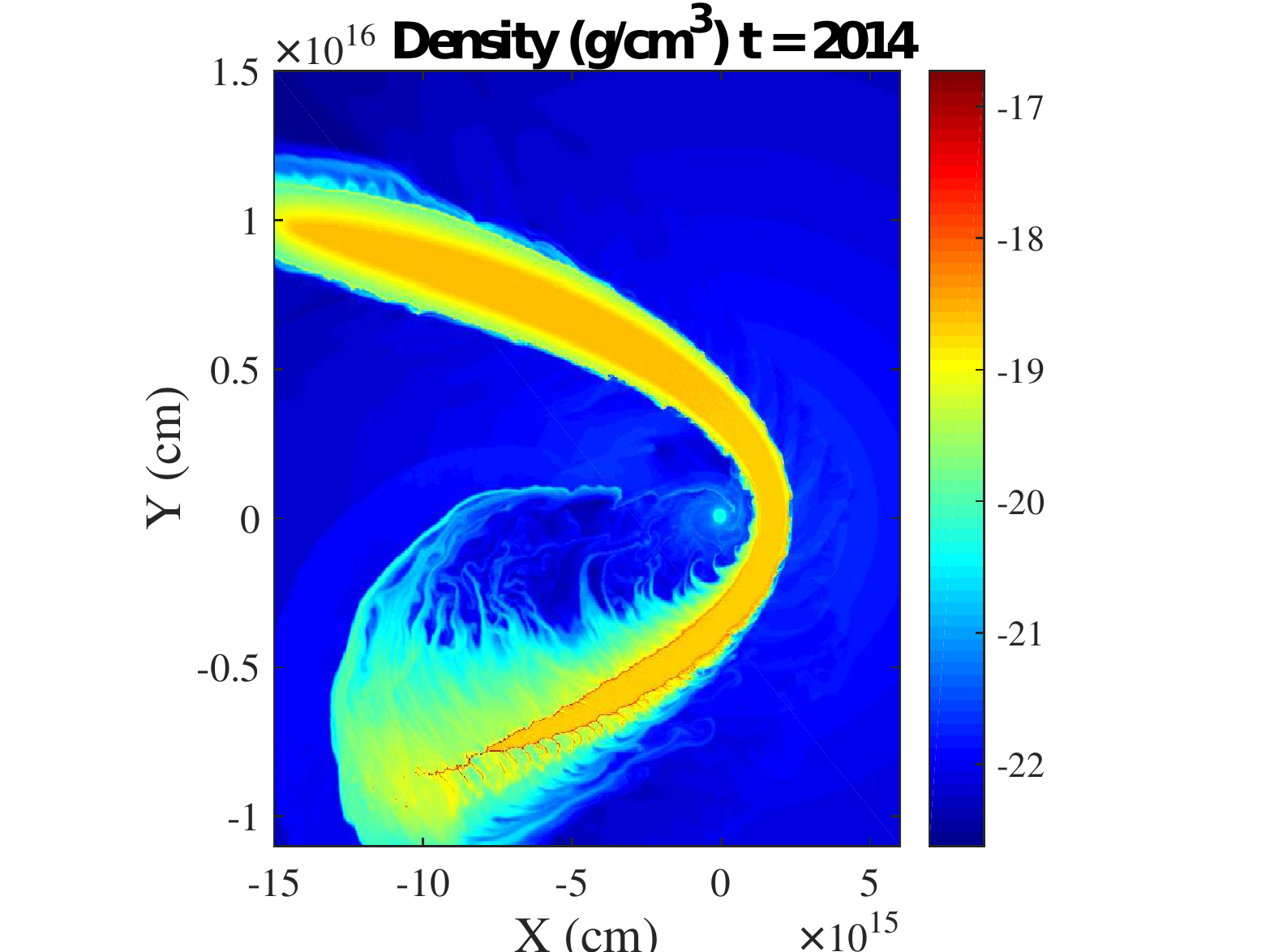}}}
	\caption{The density (log scale) of the orbital plane simulation. The left figure shows the initial set up and the right figure shows the pericenter passage. }
	\label{fig:orbit0}
\end{figure*}

In order to take into account radiative cooling, we incorporate cooling tables based on calculations performed with version 13.03 of Cloudy, last described by \cite{Cloudy}. The tables are calculated for a large combination of densities and temperatures with abundances set to GAS10 (solar composition) assuming an incident radiation field from a blackbody with a temperature of $T=3\cdot10^4$ K and an intensity of ionizing photons of $1.5\cdot10^{13}\;\mathrm{s^{-1}\;cm^{-2}}$. This intensity is a free parameter of our model and is calibrated by the requirement of matching the observed $Br_\gamma$ luminosity. 
At the end of each time step, the gas is heated/cooled by interpolating the produced tables.
Figure \ref{fig:orbit0} shows the initial conditions of the simulation as well as a snapshot taken during pericenter passage.

Preliminary runs in the orbital plane showed that the high pressure of the ambient medium near pericenter shock compress the cloud dramatically along the direction perpendicular to the orbit. This shock compression causes a cooling instability that fragments the cloud into small dense clumps. These clumps have a small surface area and hence have a reduced ionization rate from the background UV flux from nearby young massive stars. This results in a large reduction (order of magnitude) of the emitted $Br_\gamma$ emission. So far, no models using the above density profile for th ambient medium has successfully explained the observations. Therefore, in the following simulations we reduce the density of the ambient medium by a factor of 20 (found by trial and error) while keeping the temperature fixed. This corresponds to $\eta\approx40$.

\begin{figure}
\centering
\includegraphics[width=0.95\linewidth]{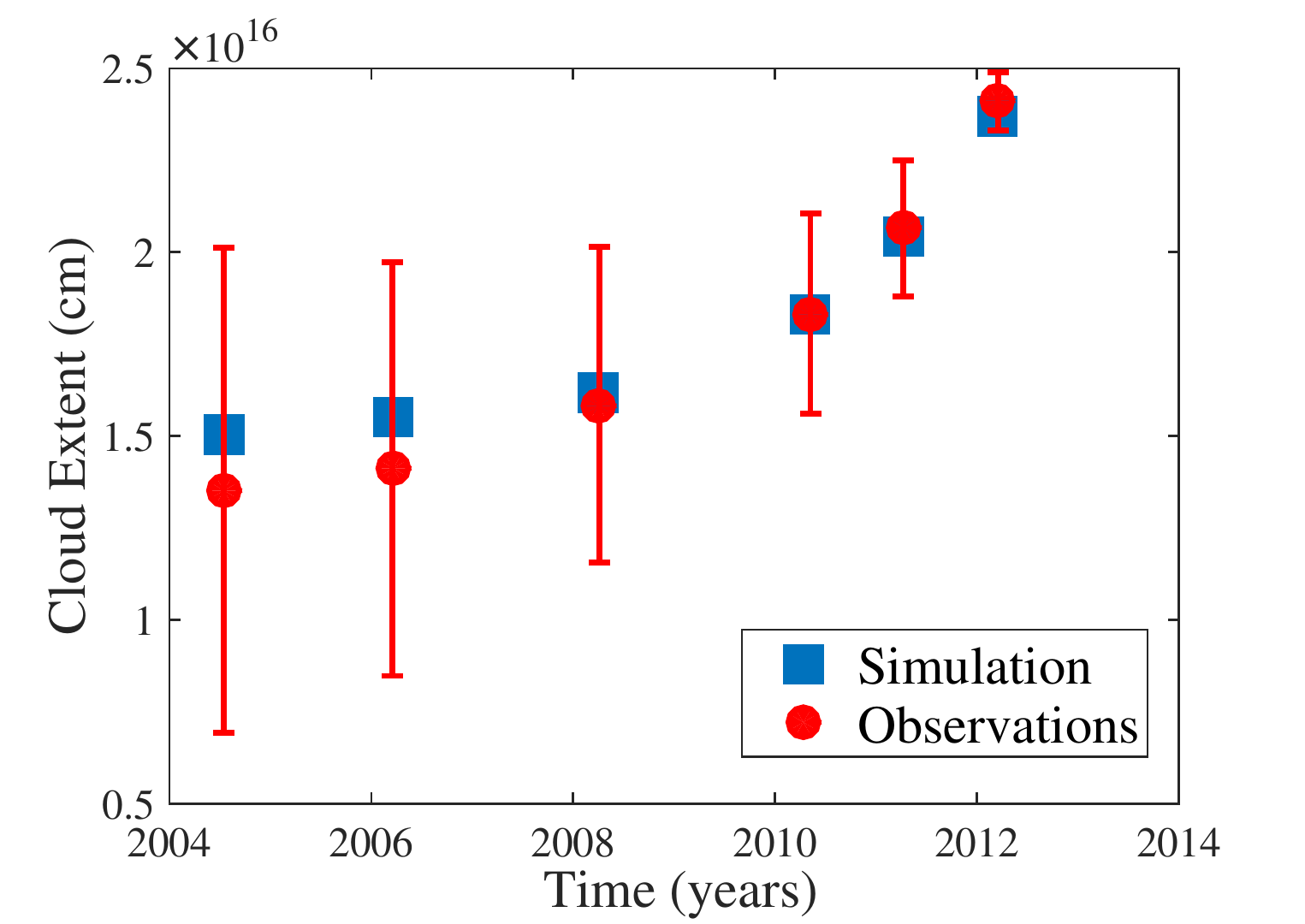}
\caption{The length of the G2 gas cloud along its orbit. The blue boxes are the measured length of the cloud from the orbital plane simulation while the red circles are the length of the cloud derived from the spread in the line of sight velocities of the $Br_\gamma$ emission, assuming tidal stretching, with a 70 km/s error in the velocity spread.}
\label{fig:g2extent}
\end{figure}

Our assumption that the cloud was already elongated in 2001 is further justified by comparing the result of our simulation with the observed PV diagrams. Assuming that the observed spread in the line of sight velocity arises from tidal stretching, the spread can be converted to the length of the cloud along the orbit. Figure \ref{fig:g2extent} compares the observed data with the measured lengths from our simulation. It is evident that there is a good agreement between the two and that the cloud is indeed already elongated in 2001.

\subsection{Vertical Collapse Simulation}
In order to gain insight about what happens in the $z$ axis during pericenter, we preform an isolated run mimicking pericenter passage.
We simulate the passage of the cloud from a distance of $2r_p$ through pericenter and until it reaches once again a distance of $2r_p$, assuming it's on an unperturbed elliptical orbit.

\begin{figure*}
	\centering
	\subfloat{{\includegraphics[width=0.5\linewidth]{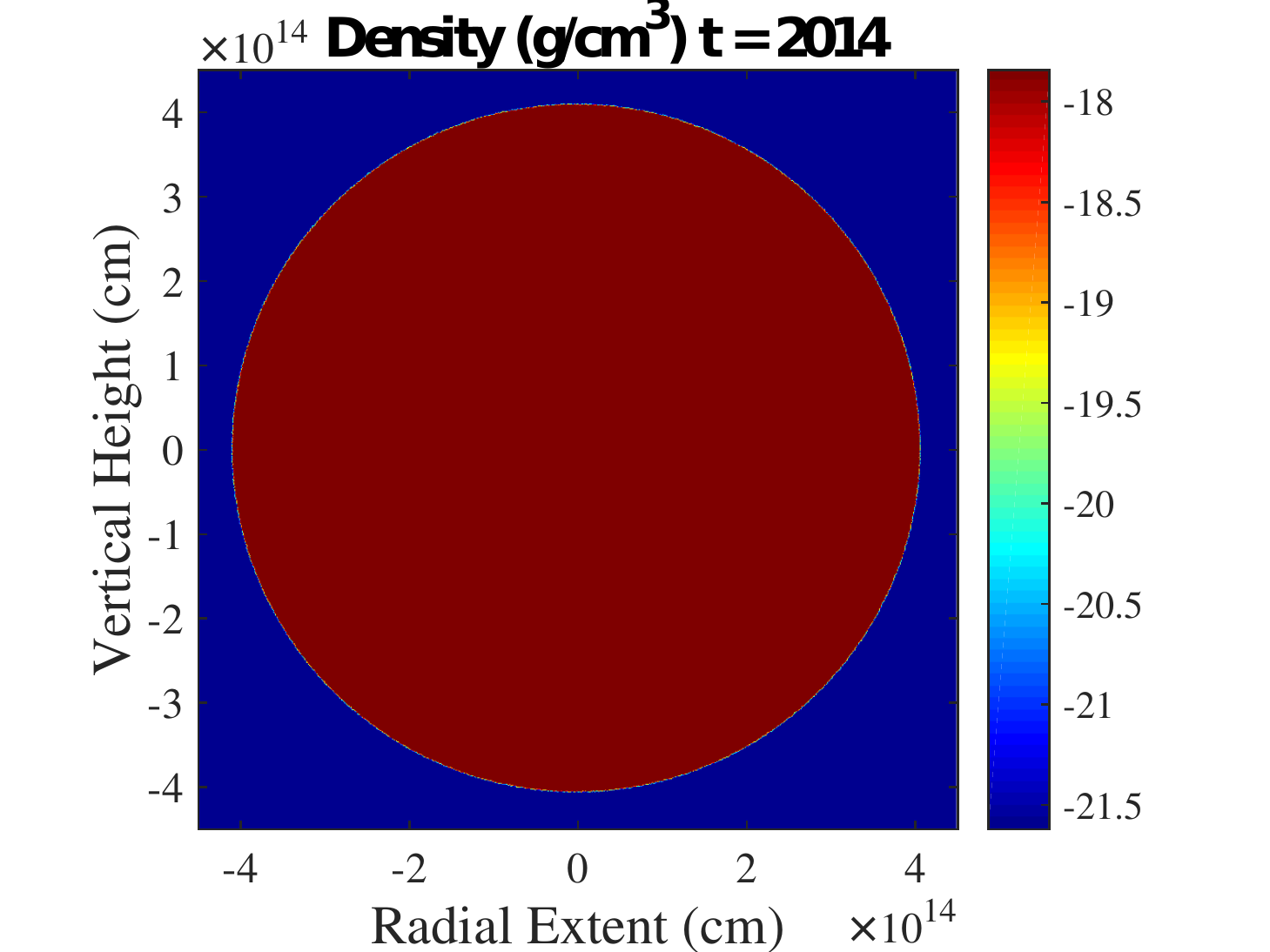}}}
	\subfloat{{\includegraphics[width=0.5\linewidth]{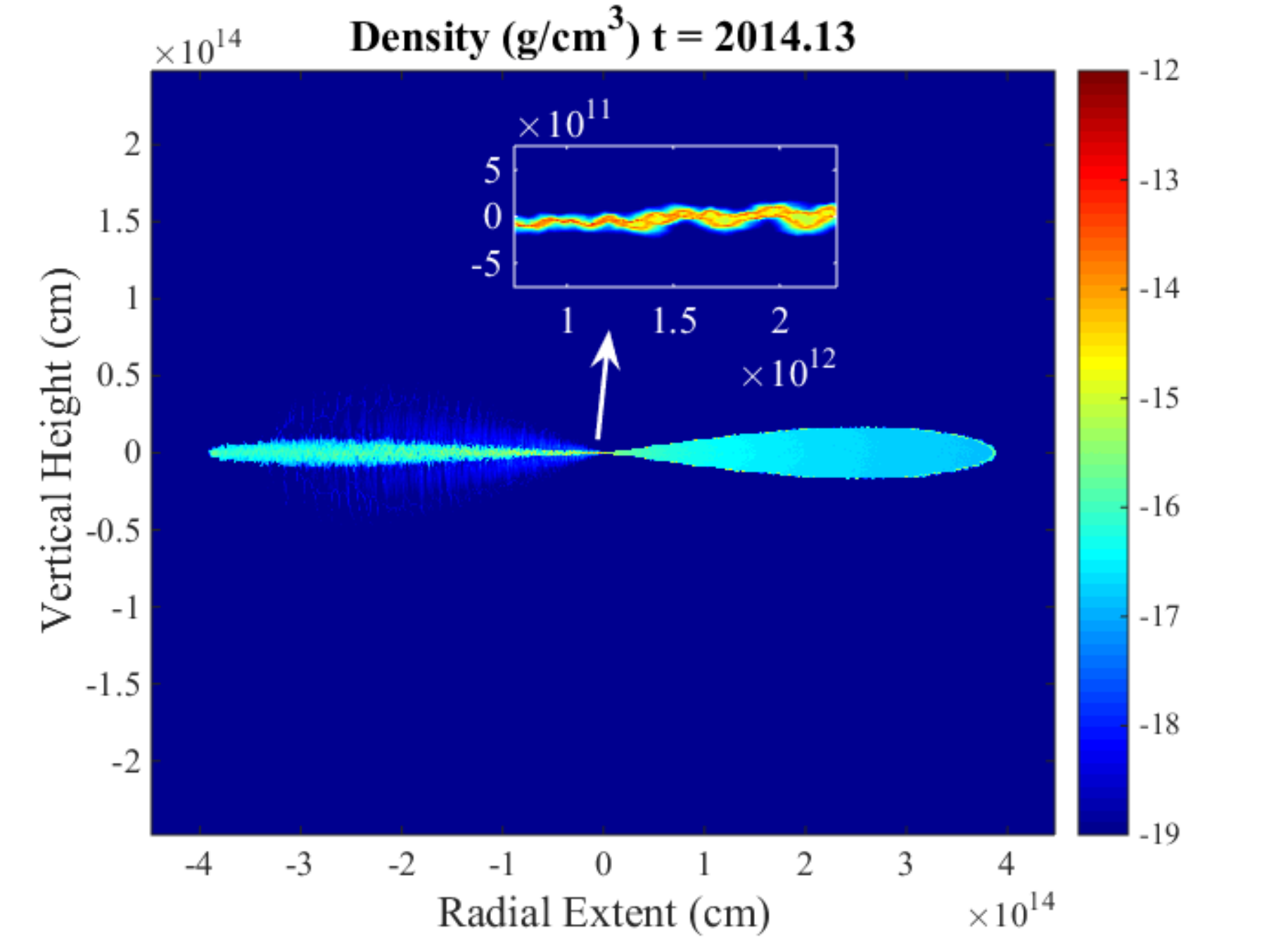}}}
	\caption{The density (log scale) of the vertical collapse simulation. The left figure shows the initial set up and the right figure shows the maximum compression achieved at pericenter. }
	\label{fig:vertical}
\end{figure*}

The cloud is set up as an homogeneous disc, representing a ``slice" of the cloud in the $r-z$ plane, with a density of $\rho = \rho_c \cdot \sqrt{r_0/2r_p} = 1.45\cdot 10^{-18} \;g/cm^3$, temperature of $10^4$ K and a radius of $R=R_c\sqrt{r_0/2r_p} = 4\cdot 10^{14}$ cm. The cloud is given a vertical velocity, $v_{z}$, that corresponds to the homologous velocity profile that is bestowed by the tidal field whose magnitude at the edge of the cloud  given is given in eq. \ref{eq:tide_vel}. 
The center of the coordinate system is centered at the center of the cloud, and we simulate the motion in the $z-r$ plane, where $r$ is the direction radial from the SMBH. Only the $z$ component of the SMBH's gravity is simulated and its magnitude is calculated from the time of the simulation which translates to a position on the elliptical orbit. 
As in the orbital plane simulation, ideal gas is assumed and the gas is cooled at the end of each time step. We simulate a box of size $[-4.5,-4.5]\times[4.5,4.5]\cdot10^{14}$ cm and we surround the cloud with an ambient gas whose properties are the same as the ones given in the orbital plane simulation. The movement along the orbital plane is imitated, therefore, by modifying the tidal field and the ambient atmosphere properties with time, according to the varying position of the ``slice" along the orbital plane. The boundary conditions are set to allow matter to flow out but not inside the domain. A total of $5\cdot10^5$ evenly spaced cells are used which give a maximum resolution of $7\cdot10^9$ cm during the collapse.
Figure \ref{fig:vertical} shows the initial setup of the cloud.

As the cloud collapses in the vertical direction due to tides, a shock wave forms at the outer parts of the cloud from the over pressured ambient atmosphere. This shock wave breaks the homologous nature of the collapse, shock heats the gas and leads to a finite height of $\approx 5\cdot 10^{10}$ cm. A snapshot through this process is shown in the right panel of fig. \ref{fig:vertical}. The left hand side of the cloud has already passed pericenter and is in the process of expanding. The central region (enlarged in the inset) is undergoing pericenter passage, while the right hand side is being compressed as it approaches pericenter.

Since the size of the cloud in the radial direction is not much smaller than $r_p$, the collapse does not happen at the same time for all of the mass elements but rather starts from the innermost mass element and travels outwards. 
After maximum compression is achieved, the cloud bounces back and expands vertically with a velocity of the order of its sound speed, which is initially very hot ($\approx10^7$K), and then rapidly cools to $10^4$K (on the order of a few $10^3$ seconds).

\subsection{Model Caveats}
Our modeling of the dynamics of the G2 gas cloud was done with two 2D simulations. In reality, the evolution of the cloud is an intrinsic 3D problem and the vertical collapse influences the dynamics in the orbital plane and vice versa.
However, at pericenter, the ratio between the density of the ambient medium and the cloud's density is less than one percent. This suggests that increasing the density of the cloud in the orbital plane, should not affect the dynamics.

Additionally, the collapse time of the cloud's vertical axis is much shorter than the time it takes the cloud to pass pericenter, so it is a good approximation to treat each part of the cloud as collapsing independently.

\section{$B$\lowercase{$r_\gamma$} Emission}
\label{sec:brg}
Once we have a handle on how the vertical scale of the cloud evolves, we can compute its expected $Br_\gamma$ emission. First, we combine the orbital plane and vertical collapse simulations to obtain the 3D structure of the cloud. The vertical height of the n-th cell in our orbital plane simulation, $z_{n}$, is assumed to be  $z_{n}=R_c\sqrt{r_n/r_0}$ until the cloud reaches $r_n=2r_p$. During pericenter passage $(r_n<2r_p)$, the height of the cloud is calculated from the vertical collapse simulation in the following manner.
Each snapshot of the vertical collapse simulation corresponds to a different radial line in the orbital plane. This allows us to calculate, for a given snapshot the height of the cloud as a function of its distance from the SMBH. By combining radial lines from all of our snapshots we estimate the height of a cell based on its location in the orbital plane.

After the cloud exits pericenter passage $(r_n>2r_p)$, we assume each cell continues to expand with a constant sound speed that corresponds to a gas with a temperature of $10^4$ K. The height of each cell is taken into account both when calculating the volume of a cell for the Br$_\gamma$ emission as well as by defining an effective density $\rho_{n,eff}=\rho_n R_c/z_n$ for all of the rest of the calculations in this section.

Once we know the effective volume of each cell, we compute the $Br_\gamma$ emission from the emissivity that is given in \cite{OF2006, Schartmann2012}
\begin{equation}
	j_{Br_\gamma}=3.44\cdot10^{-27}\left(\frac{T}{10^4\;K}\right)^{-1.09}n_pn_e\;\mathrm{erg\;cm^3s^{-1}}
\label{eq:brg}\end{equation}
where $n_p$ is the proton number density and $n_e$ is the electron number density.

A crucial aspect in calculating the emitted flux with eq. \ref{eq:brg} is correctly estimating the ionization fraction. If the ionization fraction is smaller than unity, the Brg emission will be suppressed.
There are 3 different sources for ionizations.

The first is the background UV flux that comes from young massive stars. \cite{UVnew} have estimated the flux of ionizing photons in the central parsec to be $\sim4.2\cdot10^{12}\;\mathrm{s^{-1}cm^{-2}}$.  However, we find that a better fit to the observations of the B$_\gamma$ emission requires a flux of $\sim1.5\cdot 10^{13}\;\mathrm{s^{-1}cm^{-2}}$. There is no discrepancy between the two UV fluxes, since G2's distance to the SMBH is at least an order of magnitude smaller then the location of the observational constraints. For this reason we adopt a nominal value of  $\sim1.5\cdot 10^{13}\;\mathrm{s^{-1}cm^{-2}}$ ionizing photons as the background UV flux. This is also taken into account when calculating the cooling tables. The amount of background UV flux that each cell in our simulations absorbs is calculated with a simple radiation transfer algorithm described in appendix \ref{sec:app}.

The second source of ionization is the ionizing radiation emitted due to the cooling of the shocked outer edges of the G2 cloud. For each snapshot in our vertical collapse simulation we find the temperature and density of the shocked gas and calculate the emission rate of ionizing photons ($>13.6$ eV) using Cloudy. We assume that the number of ionizations that a single photon induces is the integer part of its energy divided by the ionization energy. Our result is insensitive to the last assumption and if we assume that each photon can at most induce a single ionization our result differs by only a few percent.

\begin{figure}
\centering
\includegraphics[width=0.9\linewidth]{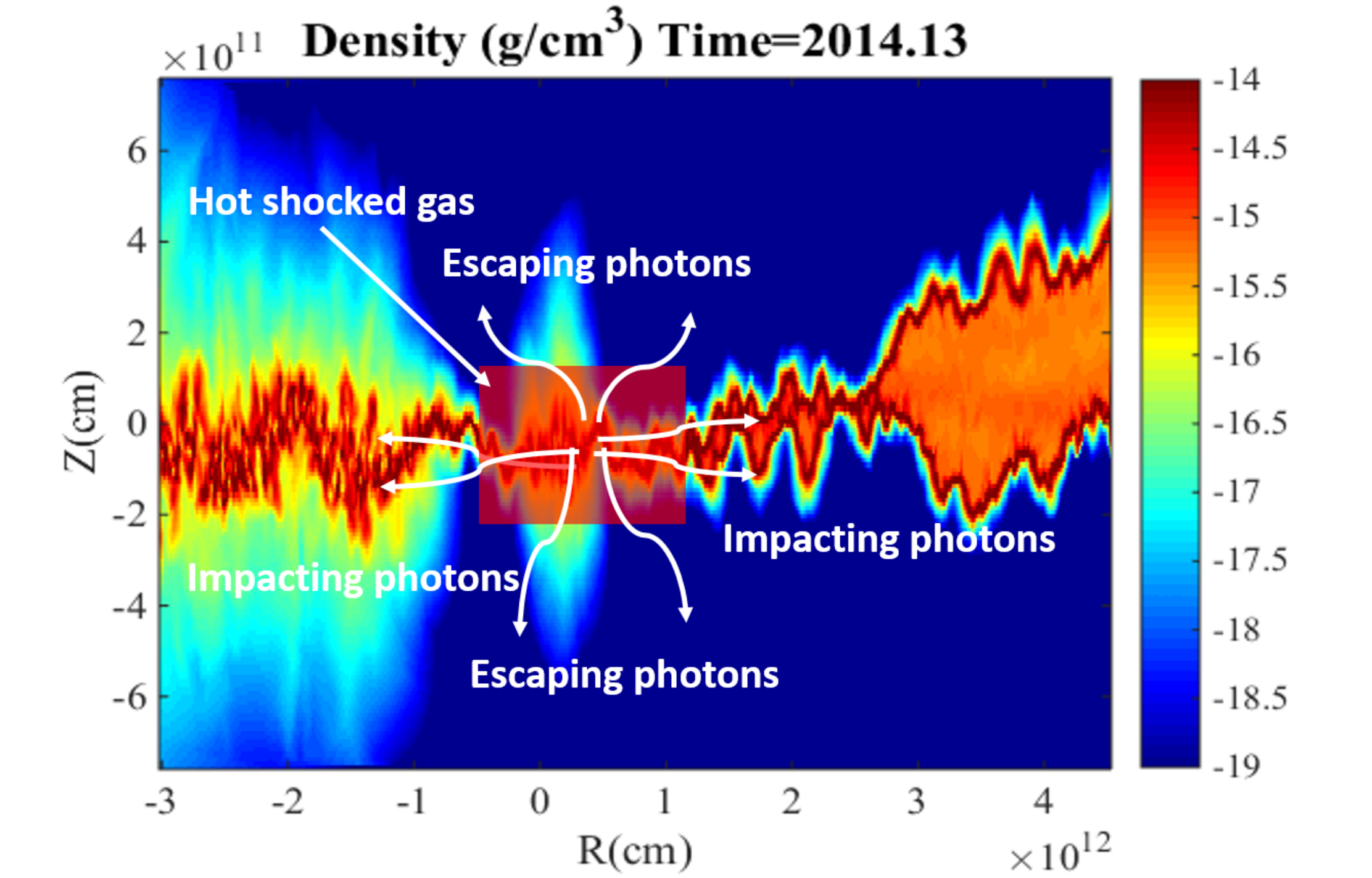}
\caption{Illustration of the emitted UV photons from the hot shocked gas during pericenter collapse. The red shaded area marks gas which is hot and did not have enough time to cool. Notice the different scales for the $x$ and $y$ axes.}
\label{fig:hotshockdiagram}
\end{figure}

During pericenter passage, the cloud undergoes an extreme vertical collapse with a velocity of $v_{col}\approx 800$ km/s. When the upper and lower halves of the cloud collide, the collapse velocity raises the temperature to $T_{shock}\approx 1.2\cdot10^7$ K. The shocked gas remains hot until it adiabatically expands vertically with a sound speed that is comparable to $v_{col}$ while simultaneously continuing to flow with its Keplerian velocity in the orbital plane.
The height of the cloud during collapse divided by the length of extent of the hot ionized gas in the orbital plane is of the order of $v_{Keplerian}/v_{col}$. Only a small fraction (the solid angle fraction is $\approx1/20$) of the ionizing photons that are emitted during the cooling of the hot collapsed cloud are therefore intercepted by the cloud. Figure \ref{fig:hotshockdiagram} shows an illustration describing the geometry of the emitted and intercepted UV photons.

The third source for ionization is shock ionization of the mass passing through the shock front at the outer edges of the cloud. As each fluid element passes through the shock front it gets shock heated and subsequently ionized. 

We find that the background UV flux from massive stars dominates the total ionizations budget at all times.

\begin{figure}
	\centering
	\includegraphics[width=0.95\linewidth]{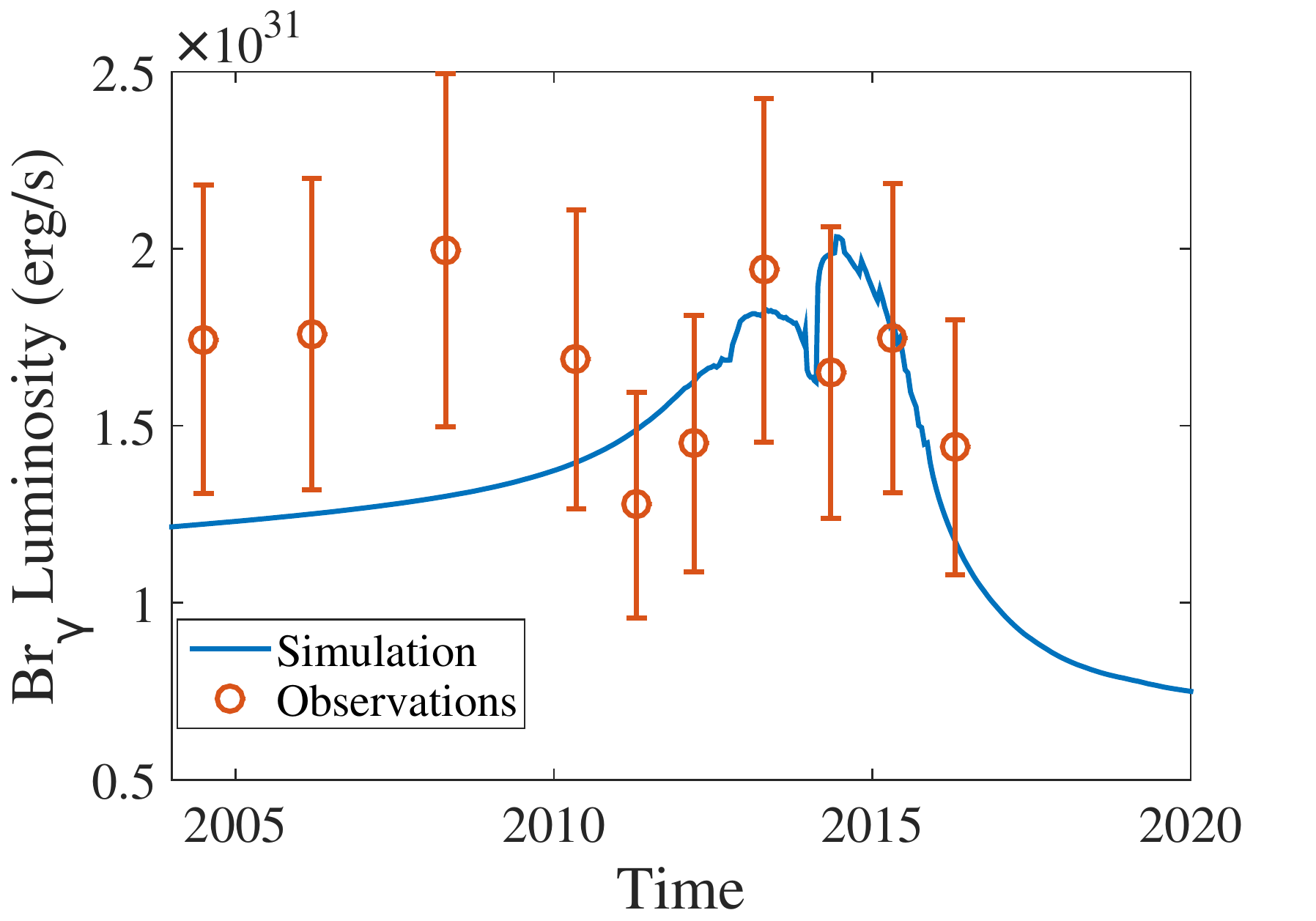}
	\caption{The integrated $Br_\gamma$ luminosity as a function of time. The solid blue line results from our simulations while the points are the observed data taken from \citet{Pfuhl} with a $1-\sigma$ error bar.}
	\label{fig:brg}
\end{figure}

For each cell we calculate the minimum between the absorbed ionizing photons rate and the emission rate of $Br_\gamma$ assuming the gas is fully ionized.
The integrated $Br_\gamma$ emission as a function of time is shown in fig. \ref{fig:brg} along with data from observations taken from \cite{Pfuhl}.

As the cloud travels towards the SMBH, its volume decreases due to tides and hence it's recombination rate and total emission increase. During pericenter passage, gas that passes through pericenter has its density greatly increased due to the vertical collapse. 
While the recombination rate for a fully ionized gas greatly increases during pericenter passage, the fact that the cloud is UV starved and that its total ionization rate is regulated by the background UV flux limits the increase in the $Br_\gamma$ emission to a mere doubling.
This translates to roughly a doubling of the emission which is now limited by the UV flux.
After pericenter passage, the emission decreases as the surface area of the cloud decreases due to the evolution of the cloud under tides.

Our simulations also allow us to create mock position velocity (PV) diagrams that can be compared with observations. We create mock data cubes in the same manner as in \cite{Schartmann2015} using pixels of size 12.5 mas and 70 km/s.

\begin{figure*}
\centering
\includegraphics[width=0.95\linewidth]{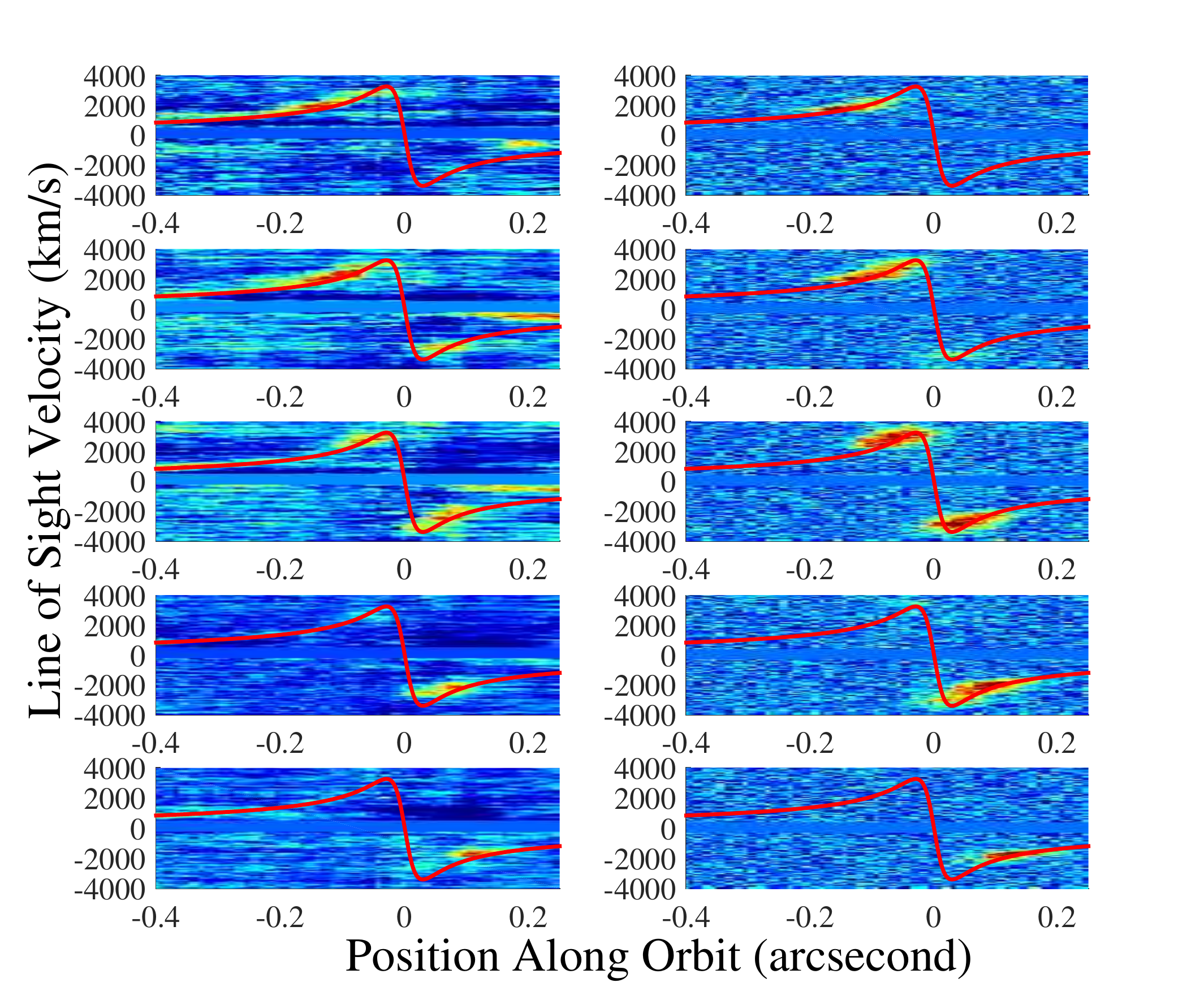}
\caption{Position Velocity (PV) diagrams of the G2 cloud. The left column is the observational data taken from \protect\cite{Plewa} while the right column are mock observations created from the simulations. The plots represent the epochs 2012.2, 2013.4, 2014.4, 2015.3 and 2016.4.}
\label{fig:fitscombined}
\end{figure*}

Figure \ref{fig:fitscombined} shows the PV diagram for several different epochs before, during and after pericenter passage. 
There is good agreement between the observations before pericenter passage and our simulations. 

\section{Discussion}
\label{sec:discuss}
We have presented a detailed model combining analytical arguments along with numerical simulations that successfully reproduce the observed $Br_\gamma$ emission. By utilizing the reduced diffusion of the moving mesh scheme along with high resolution and actively cooling the gas we were able to simulate previously unnoticed physical phenomena.
The tidal force combined with the efficient cooling collapses the vertical axis of the cloud during pericenter passage. Although the collapse produces a significant amount of ionizing radiation, due to the thinness of the cloud, most of the ionizing photons escape and only a small fraction of them end up ionizing neutral gas. The $Br_\gamma$ emission is regulated by the external UV flux from nearby massive stars. We find that the emission is UV starved, most of the gas is neutral during pericenter passage. This is true even for our enhanced ionizing background, which is 3.6 times larger than the \cite{UVnew} rate. 

Motivated by the observed duration of simultaneous blueshifted and redshifted $Br_\gamma$ emission at pericenter passage, we chose to start our orbital plane simulation with a cloud that has an aspect ratio of 4, which is consistent with the observations due to viewing angles. If the elongated cloud is integrated backwards in time under the influence of tides, it reaches a spherical shape roughly at apocenter. This strengthens the gas cloud model for the G2 object.

One of the important things we can learn from G2 is the nature of the gas in the innermost part of our Galactic Center. Extrapolating the density profile of \cite{Yuan2003} down to pericenter, results in a density $n(r=1.7\cdot10^{15}\;\mathrm{cm})_{atm}\approx 10000\;\mathrm{cm^{-3}}$. This high density corresponds to a high pressure that shock compresses the cloud during pericenter passage, leading to cooling instabilities and resulting in the fragmentation of the gas cloud. This fragmentation results in an order of magnitude reduction of the $Br_\gamma$ emission that was not observed.
In order to match the observations, the density at pericenter needs to be at least a factor of 20 lower than the extrapolated value, $n(r=1.7\cdot10^{15}\;\mathrm{cm})_{atm}\lesssim 500\;\mathrm{cm^{-3}}$. This value is consistent with the upper limit set by the lack of evolution of G2's orbital elements during pericenter passage \citep{Plewa}.

It has been proposed that there is a connection between G2 and another gas cloud, G1 \citep{Pfuhl}, that orbits the Galactic Center. 
Recent calculations \citep{Pfuhl,drag1,drag2} have shown that if a drag force is applied on G2, its resulting orbit after pericenter passage resembles the observed orbit of G1 (with a time delay). 
Since the magnitude of the drag force depends linearly on the density of the ambient medium, our reduced atmosphere results in a reduction by a factor of 20 in the drag force. 
We find that during pericenter passage the cloud's orbital energy and angular momentum change by less than one percent, which is significantly less than the difference between the observed properties of G1 and G2 \citep{Pfuhl}.
However, if G1's initial orbit was similar to G2's and its density was an order of magnitude lower than G2's, it would have experienced a drag force with the right magnitude to explain its current orbital elements.

A dilute ambient medium would also have a large impact on the expected radio emission. Assuming that the magnetic field is proportional to the gas pressure, \cite{radio_emit} calculated the predicted radio flux from the bow shock that is formed at the head of the cloud. Reducing the density of the ambient medium reduces its magnetic field (for a fixed temperature) as well as lowers the number of emitting electrons. Using equations 9 and 14 from \cite{radio_emit}, we find that a reduction by a factor of 20 in the density of the ambient medium reduces the expected radio emission by a factor of $\approx220$. This low value for the radio flux is below the current detection limit and explains why it has not been detected.

If the cloud has an initial random magnetic field that is dynamically relevant, it might prevent the vertical collapse. Equating the pressure gradient required to overcome tidal forces at pericenter yields a magnetic field of a few Gauss. 
However, since the area of the cloud in the orbital plane increases by a factor of a few during pericenter passage, the lack of compression in the vertical axis leads to a reduction in the $Br_\gamma$ luminosity as opposed to the observed increase.

In the merger model, the G2 object is a $\sim2M_\odot$ stellar object that resulted from a merger of two main sequence stars \citep{gas_star, Witzel}, and has a radius of a few AU. The $L'$ band emission of the merged star arises from blackbody radiation of warm dust emanating with a size of a few AU and the $Br_\gamma$ emission comes from an extended gas envelope that surrounds the star up to a radius of $\approx2\cdot10^{15}$ cm. 
The star's sphere of influence, defined as the region where gas would be bound to it rather than to the SMBH, is of order one AU at pericenter and grows linearly with the distance from the SMBH. 
Our simulations did not include a central object and assumed that G2 is a pure gas cloud. However, since the star's sphere of influence is small compared to the size of our simulated gas cloud, the presence of a central object is not likely to drastically affect our orbital plane simulation and its results should be qualitatively the same. 
As for the vertical collapse simulation, any gas that has a projected distance on the orbital plane of more than one AU from the star's center would behave the same as in our simulation. 
Only gas elements within a confined cylinder with a radius of an AU would be halted once they reach the star's outer radius, resulting in a compression factor of $\sim100$. Overall the resulting $Br_\gamma$ lightcurve and PV diagrams would be similar, since a stellar object has a very limited influence on an extended object such as the gas component of G2.

While we have shown that the gas cloud undergoes a drastic compression in the vertical axis, it has marginal impact on the emitted $Br_\gamma$ radiation since the majority of self-produced ionizing photons escape the system.

In general, the gas cloud model with a dilute ambient medium appears to match the observations quite well.

\section*{Acknowledgements}
E.S. and O.G. are supported by the Israeli Centers of Excellence (I-CORE) program (center no. 1829/12) and by the Israeli Science Foundation (ISF grant no. 857/14), R.S. is partially supported by and ISF grant and ICORE grant.

\bibliographystyle{mnras}
\bibliography{G2Iso_nobold}

\appendix
\section{Calculating the Ionization State}
\label{sec:app}
In order to calculate the ionization state of the gas, we adopt the following simple procedure. The results of the orbital plane simulation are mapped onto a 2D Cartesian grid whose cell size corresponds to our smallest cell in the orbital plane simulation. 

The 2D Cartesian grid is then converted to a 3D Cartesian grid by multiplying the 2D grid 20 times in the vertical direction and assigning the height of each cell to be the maximum height of the cloud (as described in sec. \ref{sec:brg}) divided by 20.
The density of the n-th cell is assumed to be the $\rho_{n,eff}$ for cells whose height is less than $z_n$ and zero for cells whose height is above the height of the cloud, and the temperature is taken from the orbital plane simulation. 

For each cell we compute it's recombination rate for case B recombination assuming the gas is fully ionized. This is the rate of ionizations a cell needs to remain fully ionized, and is the cell's effective ``opacity". 

We preform 10 iterations of injecting 0.1 of the background UV flux from the edges of the domain and count the number of absorbed photons in each cell by taking the minimum of the recombination rate and the remaining UV flux after subtracting the absorption in previously passed cells. The multiple iterations allow to correctly take into account radiation from several directions. 

This allows us to give an estimate for the ionization state of the gas in the $Br_\gamma$ calculation preformed is section \ref{sec:brg}.

\end{document}